# Gas Detection and Identification Using Multimodal Artificial Intelligence Based Sensor Fusion

Parag Narkhede [1*], Rahee Walambe [2*], Shruti Mandaokar [1†], Pulkit Chandel [1†], Ketan Kotecha [2], George Ghinea [3]

[1]Symbiosis Institute of Technology, Symbiosis International (Deemed University), Pune 412115, India
[2]Symbiosis Centre of Applied Artificial Intelligence, Symbiosis International (Deemed University), Pune 412115, India
[3]College of Engineering, Design and Physical Sciences, Brunel University, London UB8 3PH, UK
*Authors to whom correspondence should be addressed.
PN: parag.narkhede@sitpune.edu.in RW: rahee.walambe@scaai.siu.edu.in
†These authors contributed equally to this work.

## Abstract

With the rapid industrialization and technological advancements, innovative engineering technologies which are cost effective, faster and easier to implement are essential. One such area of concern is the rising number of accidents happening due to gas leaks at coal mines, chemical industries, home appliances etc. In this paper we propose a novel approach to detect and identify the gaseous emissions using the multimodal AI fusion techniques. Most of the gases and their fumes are colorless, odorless, and tasteless, thereby challenging our normal human senses. Sensing based on a single sensor may not be accurate, and sensor fusion is essential for robust and reliable detection in several real-world applications. We manually collected 6400 gas samples (1600 samples per class for four classes) using two specific sensors: the 7-semiconductor gas sensors array, and a thermal camera. The early fusion method of multimodal AI, is applied The network architecture consists of a feature extraction module for individual modality, which is then fused using a merged layer followed by a dense layer, which provides a single output for identifying the gas. We obtained the testing accuracy of 96% (for fused model) as opposed to individual model accuracies of 82% (based on Gas Sensor data using LSTM) and 93% (based on thermal images data using CNN model). Results demonstrate that the fusion of multiple sensors and modalities outperforms the outcome of a single sensor.

# 1. Introduction

Engineering innovation refers to the solving the social and industrial problems via use of the innovative engineering technologies and approaches. With the rise of industrialization and bridging of socio-economic gap between different strata of society, use of chemicals has been on rise. Assistive technology is the technological domain consisting of systems having either software or hardware alone or both designed to enhance and maintain human capabilities in situations that require special attention. Different solutions in assistive technology range from unmanned vehicles-based surveillance applications to healthcare applications like automated wheelchairs, pose estimates, etc. In this work, we propose an assistive technology solution for a very relevant problem of gas detection and identification for domestic, industrial, and outside environments.

Industrial hazards can cause chemical and/or radioactive damage to the surrounding environment. With the rapid developments in the industrialization and automated chemical plants, gas leakage is a common issue. Explosions, fires, spills, leaks, and waste emissions are some of the consequences of industrial accidents [1,2]. Residential cooking and carelessness in disposing of wastes generate unnecessary fumes, are the significant reasons of fume leakages. An article presented in the media revealed that burning wood, biomass, and dung led to 326,000 of the estimated 645,000 premature deaths from outdoor air pollution, which constitutes about 50% of the total deaths due to outdoor pollution [3]. Harmful gases such as Liquid Petroleum Gas (LPG), Compressed Natural Gas (CNG), Methane, Propane, and other flammable and toxic gases, if not used carefully and adequately, may lead to accidents and, in some cases, disastrous consequences. A gas leak is an unintended crack, hole, or porosity in a joint or machinery, which excludes different fluids and gases, allowing the escape of a closed medium. In any plant or industrial setup, a gas leak test is a quality control step that must be performed before a device is set up. As a precautionary measure, gas sensors are set up near the leakage prone equipment. However, the sensors are not able to detect gas in a mixed gas environment. Sensors are also prone to and limited to their operating characteristics.

Human intervention is not always possible in leakage situations, primarily due to the hazardous nature of gases. Smoke emissions during leakages give rise to unclear vision problems, fire and smoke leakages demand the immediate evacuation of persons with mobile disability. Breathing these dangerous fumes may lead to dizziness, unconsciousness, and mass disaster if not treated properly. In the case of gas leakage in chemical factories, it can cause explosions. Therefore, detecting gas leakages and explosions within a short period is of utmost importance. Early detection of gas leakage with higher accuracy and reliability using the state-of-art techniques is an essentially required assistive technology solution. Detecting a particular gas or different gases in the mixture of gases is also challenging and requires technological attention. Existing methods of mixed gas detection methods include a way of using a Colorimetric Tape [4]. In this method, a dry material of tape reacts with the gas being emitted and leaves a special stain for different gases under consideration. The more the gas concentration, the darker the stain on the tape [5]. Gas Chromatography is another methodology that separates mixtures of gases based on differences in boiling points, polarity, and vapor pressure [6]. This method has high separation efficiency but requires a large apparatus and workforce to operate [7]. Other than the chemical methods of gas detection and the advancements in interdisciplinary technologies, various Artificial Intelligence (AI) based techniques are also reported in the literature. Different machine learning algorithms such as Logistic Regression, Random Forest, and Support Vector Machines (SVM) are proposed in the literature for gas detection [8]. However, these methods require multiple hyperparameters tuning and statistical calculation for accurate and robust gas classification. It increases the processing time, the power used, and computations [9]. Adbul Majeed [10] provided a methodology that selected top weighted features from complex datasets for improving the time complexity as well as accuracy of the machine learning models.

Khalaf [11] proposed an electronic nose system of classification and concentration estimation that uses least square regression. An array of eight different gas sensors is used to identify gases' concentration in [12]. In this work, Deep convolutional neural networks are employed for the application of gas classification. It was shown that the deep learning algorithms can learn features from the measurements from gas sensors in a better way and can achieve higher classification accuracy. Bilgera et al. [13] presented a fusion of different AI models for Gas Source Localization to determine the point of leakage in a ground using six various gas sensors. Pan et al. [14] presented a deep learning approach consisting of a hybrid framework comprised of the Convolutional Neural Network (CNN) and Long short-term memory (LSTM) to extract sequential information from transient response curves. Fast Gas Recognition algorithm based on hybrid CNN and Recurrent Neural Network (RNN) is presented in [9]. It was shown that the fusion model outperforms Support Vector Machine (SVM), Random Forest, k-nearest neighbors. Liu et al. [15] described two network structures, Deep Belief Networks and Stacked Autoencoders, to extract abstract gas features from E-nose. Then the Softmax classifiers are constructed using these features. These reported approaches use sequential methods based on the gas sensor data directly.

However, there are several issues with using only a gas sensors-based detection and identification approach. The primary reason is that the proportion of gas in air is very low in some cases, and the gases are not identifiable with standard gas sensors. This generates false negatives or false positives and hence hampers the detection accuracy of the system. Additionally, low-cost sensors are typically less sensitive and may not provide accurate measurements. Another method observed for gas detection is the use of thermal imaging. When a gas is leaked, the surrounding temperature increases compared with the normal conditions. The increase in temperature can be characterized and analyzed by thermal imaging cameras. This concept can be utilized to detect leakages [16,17]. The system for Methane and Ethane gas leak detection using a thermal camera is proposed in [18]. Jadin and Ghazali [19] presented a method for detecting gas leak using infrared image analysis. The system was designed by the technique of image processing, which are data acquisition, image preprocessing, image processing, feature extraction, and classification.

Single modality sensing methods may not achieve the system's required accuracy and robustness as such systems are limited to sensor characteristics. Individual sensors are limited to temporal and spatial characteristics [20]. A thermal imaging system can identify the presence of gas but fails to identify its type. Hence, a concept of multimodal/multi-sensor data fusion came into existence. Data fusion combines information from multiple sources to obtain the better output compared to any individual modality taken alone [20]. Kalman filter proposed in [21] is one of the most widely used sensor fusion algorithms in robotics applications like position and orientation estimation, guided vehicles, etc. However, it requires the input data from two sensors in a similar format. In the situation under consideration, the gas sensor data is a scalar value whereas input from thermal image is a two-dimensional vector. Hence, Kalman filter cannot be used in this application of fusion 1D and 2D vectors. With the advancement and flexibility of AI frameworks, a combination of different AI algorithms can be used to extract important features in an efficient and improved manner and improve classification accuracy [22,23,24]. This paper presented an AI-based methodology that employs the Deep Learning (DL) frameworks for performing a fusion of multimodality data from multiple sources to detect and classify the gasses. The system is equipped with various gas detecting sensors, and a thermal imaging camera and sensor fusion is performed using the DL algorithms.

The focus of the proposed method is to extract features using two different deep learning paradigms and apply an early fusion method to fuse these features to train a classifier for detecting and subsequently identifying the gas. The proposed method can be used to detect a particular gas in a mixed environment of gases. It does not require a manual operator to operate and is a more robust solution as it incorporates the measurements from multiple gas sensors and thermal imaging cameras. In case one modality is generating false negatives, the fusion with other modality can help identify the correct outcome more effectively. On the other hand, if one modality is giving false positives, the other modality helps to bring down the combined accuracy of fused output, thereby providing accurate predictions.

The main contributions of the paper can be listed as follows:

1. an innovative multimodal AI-based framework for the fusion of two separate modalities for robust and more reliable gas detection1 is proposed and presented
2. the use of early fusion of the outputs of deep learning architectures CNN and LSTM is demonstrated for Gas Detection and identification of the leaked gases

In summary, the main contributions of this work are twofold. Firstly, multimodal AI-based framework for the fusion for gas detection and identification is presented in this paper. This framework is faster, easier to deply and generic. Secondly, the use of early fusion of the of outputs from CNN and LSTM is demonstrated for Gas Detection and identification of the leaked gases. The vanilla architectures are considered for the implementation of CNN and LSTM frameworks. Having advanced frameworks like AlexNet [25], ResNet [26] will add to the computational complexity of the system due to very deep architectural frameworks. the use of CNN facilitates faster processing and is also suitable for the deployment in real-time systems. The results show that false positives and negatives in the fused output are lower than the individual modalities. The experimental setup is designed to collect the real-time data using a gas sensor array and thermal camera, to preprocess the collected data and validate the developed framework. Our approach is highly generic and can be extended to a number of other applications involving multiple sensors and their data fusion. Innovation lies in the development of state-of-the-art AI techniques for solving a highly relevant social and industrial issue of identifying gas leakage and controlling it in time to reduce loss of property and human lives in extreme cases.

The paper is organized as: **Section 2** provides a brief overview of AI-based multimodal fusion methods. The frameworks for data collection and preprocessing along with the proposed system architecture are presented in **Section 3**. **Section 4** provides a detailed discussion on obtained results, and **Section 5** concludes the paper by mentioning future scope.

## 2. Theoretical Background

Fusing the data from multiple sensors makes the system more robust and reliable than the single sensor-based systems. There are various methods of sensor fusion using AI paradigms proposed in the literature. This section briefly discusses these methods as a precursor to our system framework and experimentation setup.

**2.1. Methodologies for Multimodal Data Fusion**

A modality refers to something that can be experienced in the environment. It is a type of information that can be felt and is stored. Some examples could be–text information, image information, smell, taste, auditory, video, and touch. Multimodal Sensor Fusion refers to combining sensor data from different sources to produce more consistent, accurate, and useful information than individual sensors to reduce false positives and false negatives. The fusion architectures can be of three types: early fusion, late fusion, and hybrid fusion [27,28]. Early fusion combines the raw data or the features extracted from the raw data [29]. This is a suitable technique when there exists a high correlation between modalities. The feature extracting algorithms are applied to the individual modalities and then fused together using the process of concatenation to get the final feature vector. A classifier model is trained using this feature vector, and final predictions are made. In this method, the fusion is performed before the classification, which allows the interaction of features at a low level. In late fusion, the decisions are taken based on the individual modalities separately. Predictions from individual modalities are then combined the using statistical method like mean, mode, median, etc. As it is a combination of decisions, it is also known as Decision Fusion Technique. This technique is preferred when there exists a time relationship between the modalities. Hybrid fusion combines the advantages of early and late fusion for better fusion of features as well as decisions.

**2.2. Convolutional Neural Network**

Each Thermal image consists of non-linear features and are stored digitally in RGB format. Simple Neural Networks are not able to generalize complex patterns in images. Convolutional Neural Networks (CNN) learns to recognize differences and patterns in images. CNN [30] consists of - Convolution, Max Pooling, Flattening, and ANN layers. The primary purpose of convolution is to find features in an image using a feature detector and put them into a feature map.

**2.3. Recurrent Neural Network**

Recurrent Neural Networks (RNN) [31] consists of an essential memory element due to which the present output depends not only on current input but also on previous input. However, as the input sequence size increases, the problem of vanishing gradient is observed during backpropagation. This problem makes RNN unsuitable for applications requiring long-term dependencies. To overcome this, advanced versions of RNNs known as Long Short-Term Memory (LSTM), consisting of gates and memory elements, were introduced. These gates help regulate and extract information from the input and pass on gradients to the next node enabling the new sequence to be trained as equivalent as the earlier sequence and prioritize learning [32]. Also, LSTMs are more effective than conventional RNN [33].

Sensor measurements are a continuous stream of data, and hence LSTM framework is applicable for extracting the features from the sensor measurements. The thermal camera provides images, and CNN is an appropriate choice for feature extraction. The two considered modalities are having different characteristics and do not have any time-level correlation. Hence, in our proposed framework, we have employed early fusion of features extracted by the LSTM model from gas sensors and by the CNN model from the thermal images data. The further section provides the details of the pipeline for data collection using the specified sensors, preprocessing the collected data, and developing the fusion frameworks for the proposed work.

## 3. Framework for System Design and Experimentation

The system consists of gas sensors and a thermal camera for identifying the gas concentrations and thermal images of the type of gases. The block diagram indicating the data collection process is presented in Figure 1. Figure 2 provides the structure and steps followed for training the network and Figure 3 indicates the testing phase. The detailed description for the processes indicated in these figures is provided in further sections.

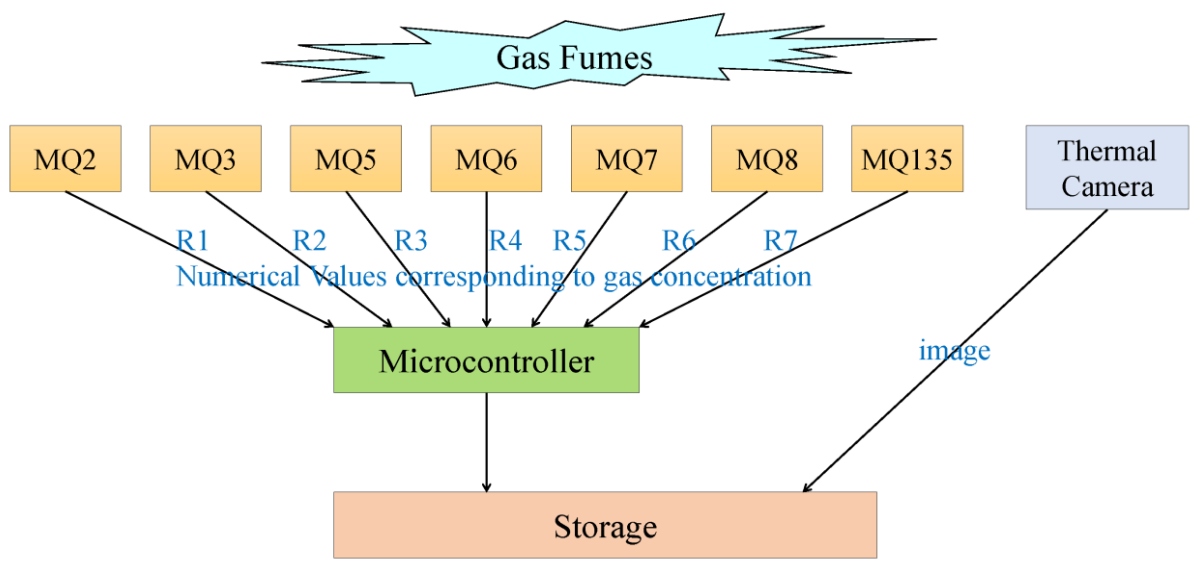

Figure 1: Process of Data Collection.

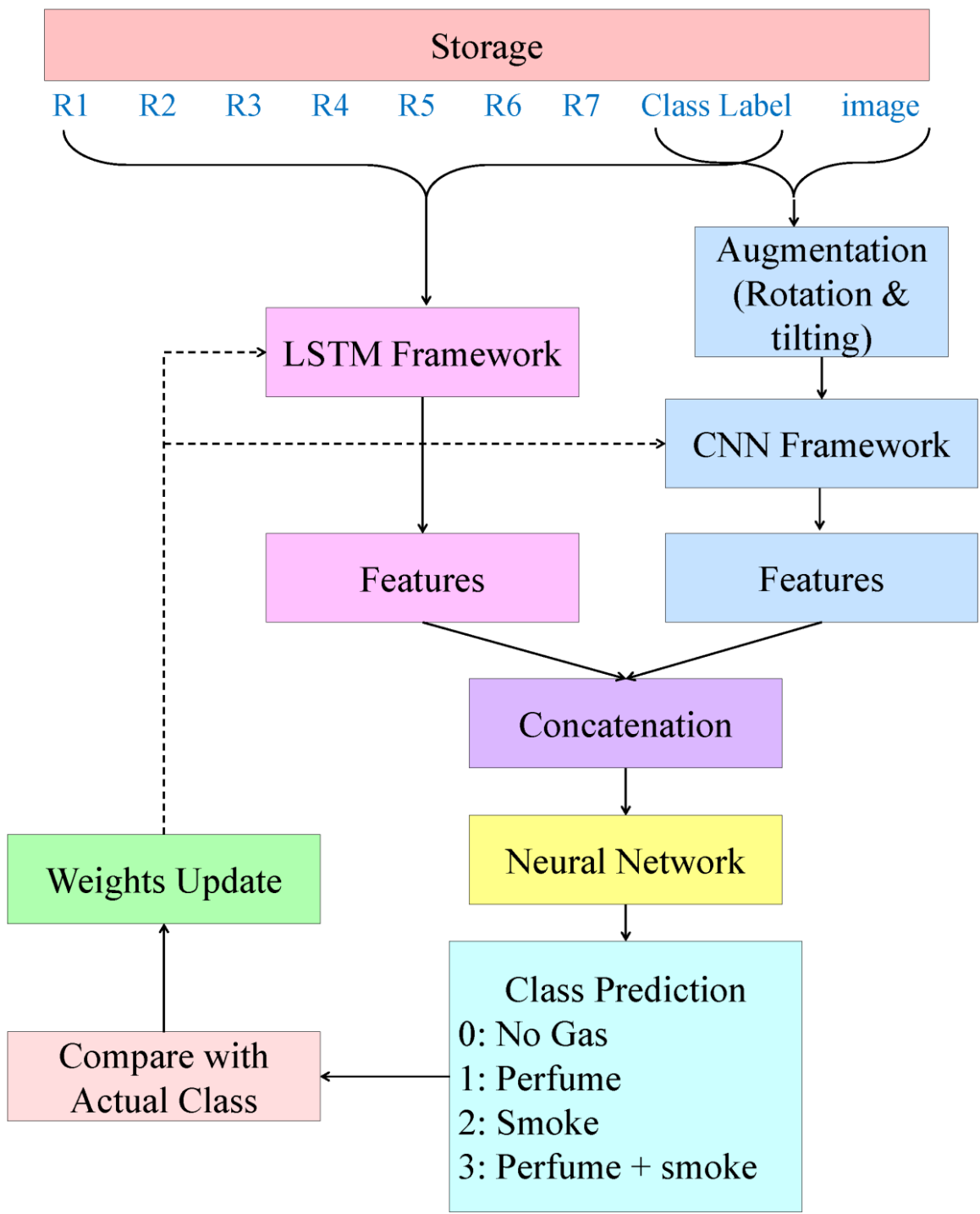

*Figure 2: Network Training Process*

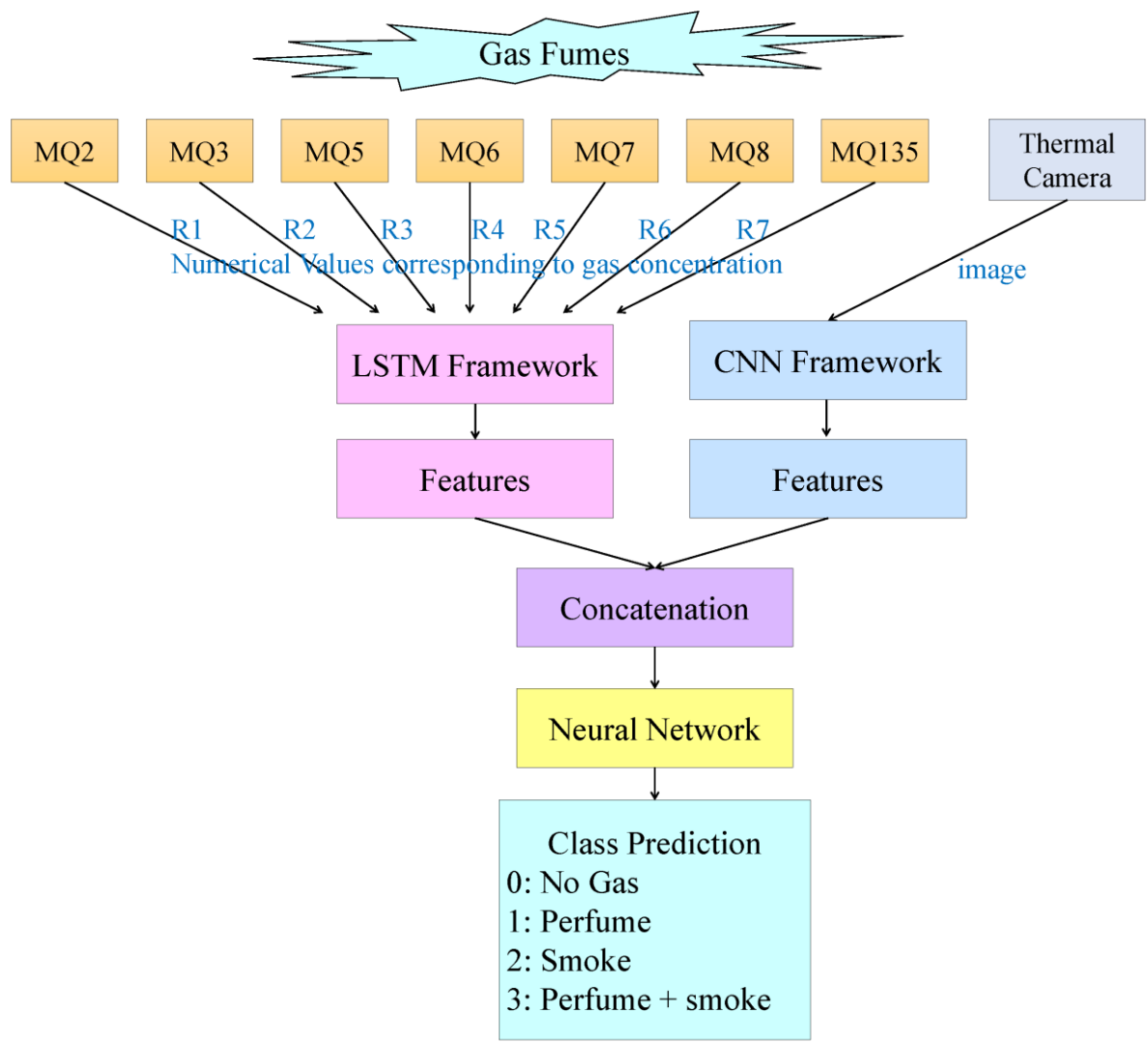

*Figure 3: Network Testing Process*

**3.1. Gas Sensors**

Gas Sensors detect the presence of gas by converting the chemical information to electrical information. Metal Oxide Semiconductor (MQ) gas sensors are appropriate as they are compact, have fast response speed, and long service life [34,35]. Each sensor consists of a heating element that produces the analog output voltage proportional to the gas concentration. The performance of Gas sensor depends on various sensor characteristics like sensitivity, selectivity, detection limit, response time, etc. [36]. Different gas sensors namely MQ2, MQ3, MQ5, MQ6, MQ7, MQ8 and MQ135 are used in the present work. These sensors are sensitive to various gases like Methane, Butane, LPG, Alcohol, Smoke, Natural Gas, Carbon Monoxide, Air Quality etc. (**Table 1**).

Table 1. Gas sensors and sensetive gases.

| Sensor | Sensitive Gas |
|---|---|
| MQ2 | Methane, Butane, LPG, Smoke |
| MQ3 | Alcohol, Ethanol, Smoke |
| MQ5 | Natural Gas, LPG |
| MQ6 | LPG, Butane Gas |
| MQ7 | Carbon Monoxide |
| MQ8 | Hydrogen Gas |
| MQ135 | Air Quality (Benzene, Smoke) |

### 3.2. Thermal Camera

Thermal camera is a device that measures the temperature variations using the infrared light. Every pixel on a camera image sensor is an infrared temperature sensor and gets a temperature of all points at the same time. The images are generated according to temperature format and displays images in the form of RGB. Unlike normal imaging cameras, thermal camera is not constrained by dark surroundings and can work with any environment regardless of its shape and texture [37]. Seek Thermal Camera, used in this work, is a compact thermal camera consisting of 206 × 156 Thermal Sensor, a 36-degree field of view, measurement of temperature range −40 °C to 330 °C, framerate <9 Hz, and 32,136 Thermal Pixels to be able to see a thermal image easily.

The gas sensors and thermal camera are used simultaneously to collect data for training and testing of the developed fusion model. The next part of the paper describes the data collection and its preprocessing in detail.

### 3.3. Data Collection and Preprocessing

To the best of the authors' knowledge, no data consisting of thermal images and gas sensors for the representation of gas has yet been collected and available in the open domain for direct use. Hence, in this work, data of the sensors and thermal imaging camera is collected manually for model training and validation purposes.

The experimental data is collected through an array of 7 gas sensors as well as using the Seek Thermal Camera. The gas sensors were placed at 1 mm apart.

In the experimentation, two specific gas sources are identified, namely, the gases originating from perfumes and gases emitted by incense sticks. The experimentation setup and workflow for the data collection is shown in **Figure 4**.

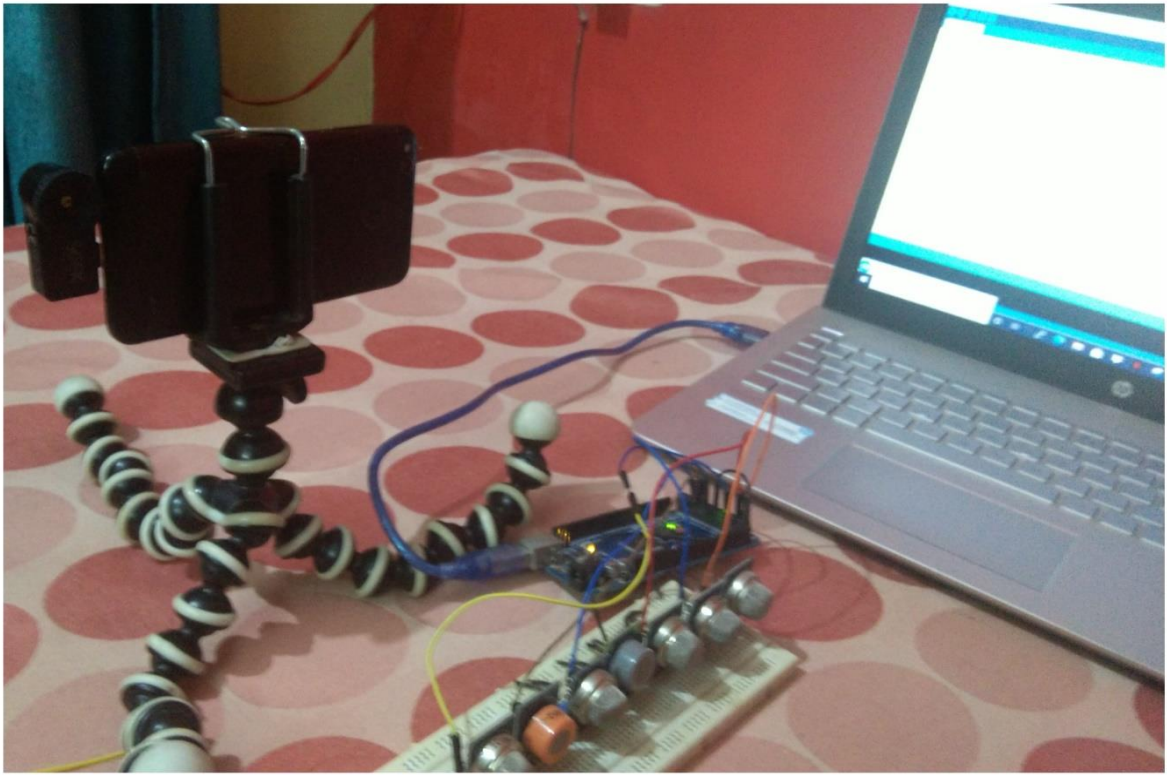

*Figure 4Experimental Setup for data collection.*

Sensor readings and thermal images are recorded for each of these two gas sources were collected at a time interval of 2 s continuously for one and a half hours. In this time, gas was sprayed with an interval of 15 s for the first 30 min, with 30 s intervals for the next 30 min and 45 s intervals for the next 30 min. A few representative samples for three classes (no gas, perfume, and smoke) with the thermal image and corresponding gas array data are shown in **Table 2**. The sensors provide the analog voltage equivalent to the gas concentration. The analog value is converted to the 10-bit digital value using an analog to digital converter. These 10 bits of digital values are shown for representation purposes in **Table 1**. Each sensor is sensitive to more than one gas, and hence sensors are calibrated appropriately. A data set in total consists of 6400 samples where 1600 samples belong to perfume, 1600 samples belong smoke, 1600 samples belong to mixture of perfume and smoke and 1600 samples belong to neutral environment (No gas).

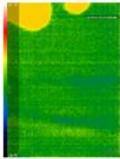

Table 2. Data samples for thermal image and their corresponding gas array obtained for 3 classes.

| | No Gas | Perfume (Alcohol) | Smoke |
|---|---|---|---|
| Thermal Image | | | |
| Gas Sensor Measurements | 558 516 376 336 665 450 415 | 808 520 515 485 692 754 513 | 550 343 371 400 572 583 304 |
| Thermal Image | | | |
| Gas Sensor Measurements | 791 520 510 455 690 733 533 | 800 521 508 481 686 746 505 | 537 354 337 374 562 547 279 |

### 3.4. Data Preprocessing

Deep learning models require a large amount of training data for appropriate and efficient operation. Due to the availability of limited data, data augmentation techniques are used, which helped to increase the dataset size. The diversity of limited thermal images is increased using data augmentation techniques such as rescaling and resizing. The **Figure 5** shows the ground truth image (**Figure 5**a) and all images generated using rotation and tilting operations (**Figure 5**b).

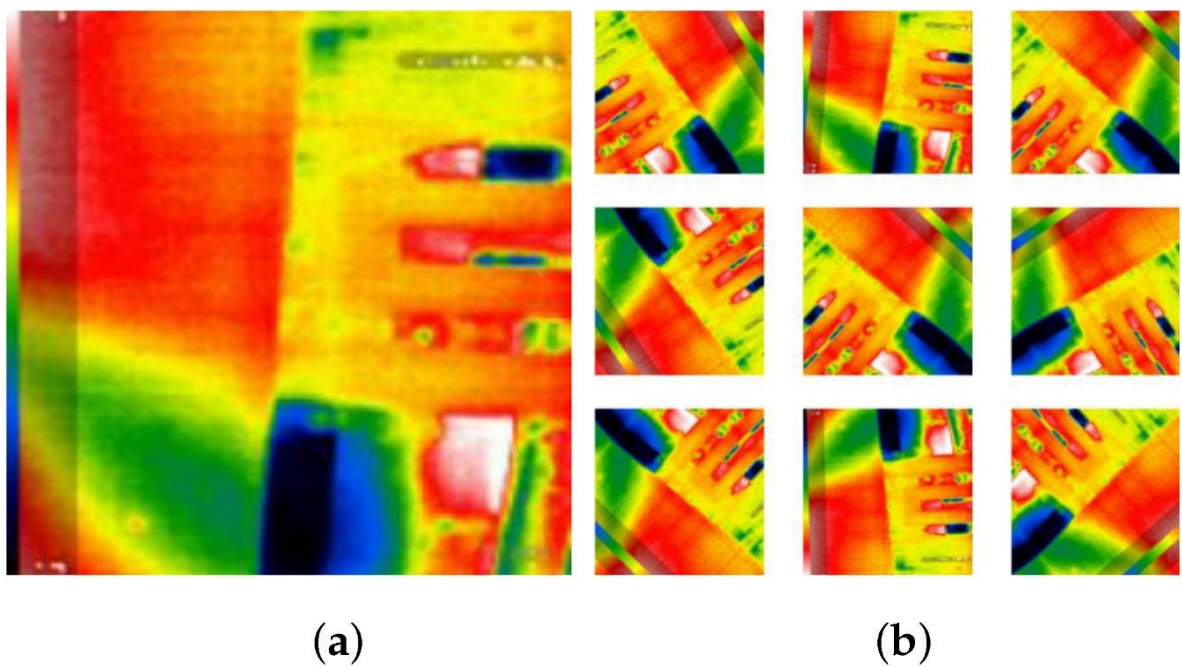

*Figure 5 Ground Truth image (a) and augmented images (b).*

### 3.5. Feature Extraction from Thermal Images Using CNN

A total of 6400 thermal images and corresponding 6400 labels (No Gas, Alcohol, Smoke, mixture of Alcohol and Smoke) are considered in this experimentation. A train-test split of 80:20 was done such that out of total images, 4096 are used for training and 1024 samples are used for Validation whereas 1280 images are used for testing purposes.

In the process of development of the CNN model, multiple experimentations were carried out with different architectures, and various hyperparameter tuning approaches were applied. It was found that three convolution-pooling layer architecture followed by a dropout layer (dropout of 0.25) is providing the best accuracy and recall. Model is optimized with different optimizers and the best performing optimizer is selected for further processes. An ADAM optimizer with a 0.001 learning rate with a decay of $1 \times 10^{-3}$ and L1-L2 regularization (0.005) in the first two Conv-Max Pool pairs are applied to avoid overfitting of the model. The model is trained for 300 epochs, which resulted in the testing accuracy of 93%.

### 3.6. Feature Extraction from Gas Sensor Measurements Using LSTM

Sensor measurements are sequential and hence sequence model namely LSTM Network is used for extracting the features from these measurements. The architecture of the LSTM model consists of the input layer followed by a single LSTM layer with 5 cells. LSTM layer is regularized with L2 regularization. The LSTM layers are followed by the classifier layer with the Softmax activation function.

This LSTM network was trained on different optimizers with a fixed learning rate of 0.001 to find the best optimizer. Through the trial and error, it was observed that Adam optimizer was fitting to the model the best and also converging quickly. Hence Adam optimizer is selected for analysis and experimentation work. It can be observed that Adam optimizer fits and converges quickly. The model is trained for 300 epochs, and we obtained the testing accuracy of 83%.

### 3.7. Multimodal Fusion of Image and Sequence Data

In this phase of the work, the features extracted from the thermal images and gas sensor measurements fused for accurate decision making. The proposed architectures of the image and sequence data fusion model are presented in **Figure 6** (early fusion) and **Figure 7** (late fusion). The focus of the work was to build a fused classifier that consists of both gas sensor sequence array and thermal images. In the fusion process, LSTM and CNN's output must be in the same feature space before fusion can be performed.

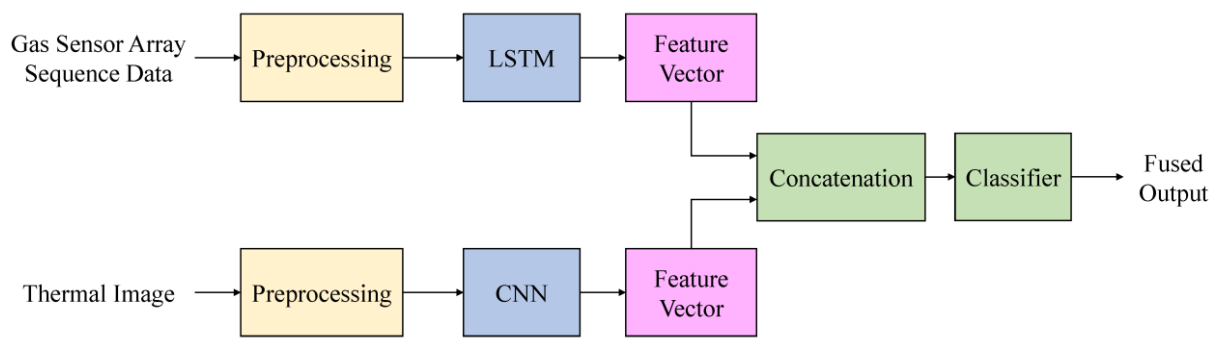

Figure 6 Framework for the proposed Early fusion model.

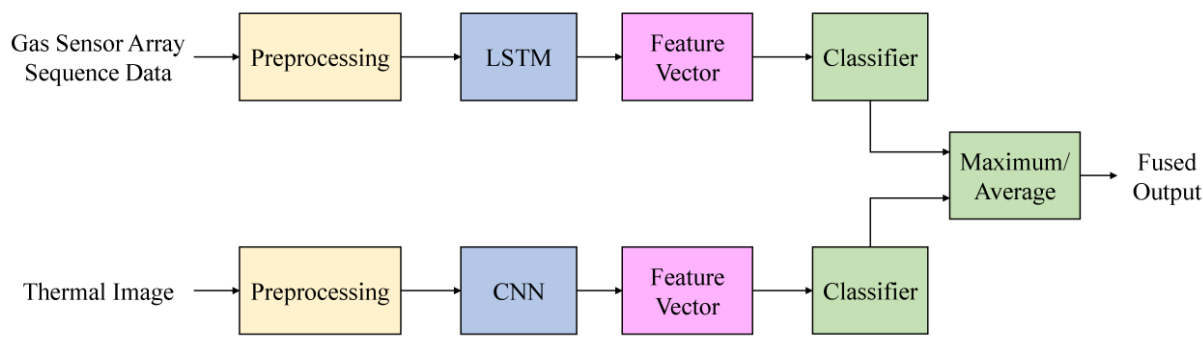

Figure 7 Framework for the proposed Late fusion model.

The fusion model is optimized with an Adam optimizer with a 0.001 learning rate and $1 \times 10^{-3}$ decay. Regularization (0.005) is applied for avoiding overfitting of the fusion model. The model is trained for 300 epochs, which resulted in the testing accuracy of 96%.

Late fusion model is also implemented for the fusion of gas sensors array data with the thermal image. Late fusion being the decision level fusion, the predictions of from individual models namely LSTM model and CNN model are obtained individually. Then the Late fusion process is applied in two ways. In first trial, maximum of the predictions from individual results is taken as final fusion value. Hereafter, this is referred as Max fusion. In another trial, arithmetic average of the individual model predictions is considered as final fusion, referred as Average fusion.

The presented models of early and late fusion are implemented and validated with the dataset available. The next section describes the results obtained and comparison between the fusion models.

## 4. Results and Discussion

The multimodal AI-based fusion model for gas detection and is presented in this work. Two modalities, namely, thermal images and gas sensor measurements, are considered in this work of gas detection. The CNN architecture is applied for extracting features from the thermal images, whereas, LSTM framework is used for extracting features from the sequences of gas sensor measurements. The implementation of the proposed model in done using the Python 3 using Keras framwork on TensorFlow platform. Open source Google Colab GPU is used for training and testing of the proposed model. It is based on Intel Xeon Processor with 13 GB RAM. The CNN model starts converging at around the 20th epoch, whereas LSTM reaches convergence at around the 90th epoch. It was observed that the fused model stabilizes at around the 20th epoch itself. The accuracy of the gas sensor array is comparatively lower since the outcome of one sensor (out of 7 sensors considered) is typically not very accurate due to the mixing of gases in the air. The thermal camera-based model individually performs comparatively better; however, in the air, the thermal signature of gaseous emissions may be generated due to multiple gases or multiple sources of exhausts, and having a gas sensor to validate the type of gas is extremely helpful in identification. It was noticed and observed that the individual models are underperforming compared to the fusion models. In the fusion models, the individual modalities either collaborate or oppose the outcomes of the individual modality, thereby making the system more reliable and accurate. By performing regularization techniques on individual models, namely CNN and LSTM, testing accuracy of 93% with Thermal Images and 82% for Gas Sequences is achieved. However, the early fusion of features from both CNN and LSTM has provided the testing accuracy of 96%, which is greater than individual models accuracies. In the case of late fusion (max fusion and average fusion) the accuracy was observed to be around 96%.

**Table 3** shows the individual training and testing accuracy, loss, precision, recall, and F1 scores for all four classes considered in this study. The accuracy comparison for the individual models is shown in **Figure 8**. It can be observed that the fusion models outperform the individual models as the predictions in the individual models are based on both the modality data. It can also be noticed that the accuracy of early fusion is slightly higher than the late fusion models as in this case the fusion happens at the feature level which allows the interaction amongst the modalities. The confusion matrices are also plotted for all the frameworks and are shown in **Figure 9**.

## ACCURACY COMPARISON

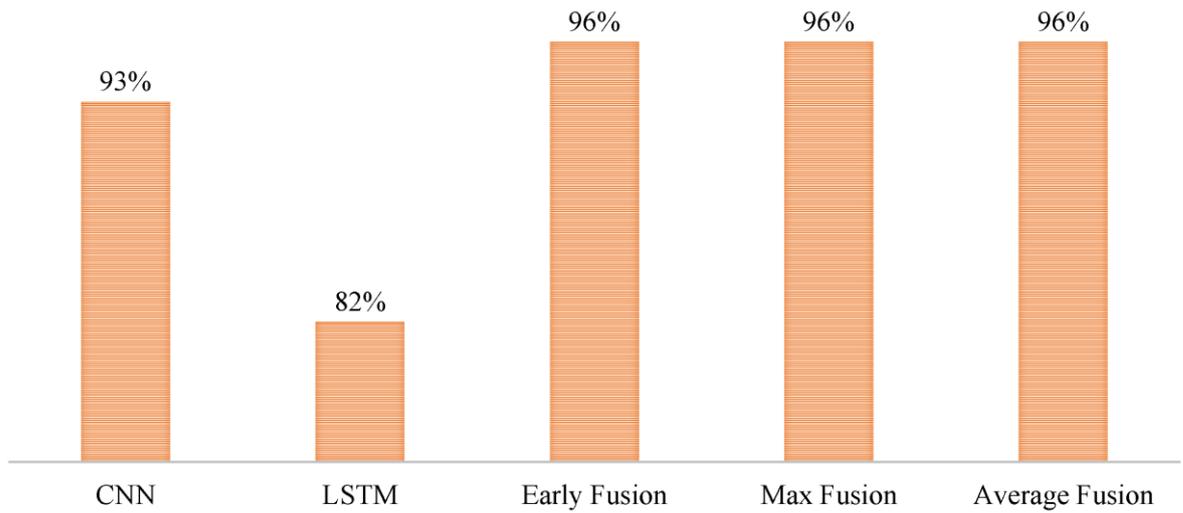

*Figure 8 Accuracy comparison of different models.*

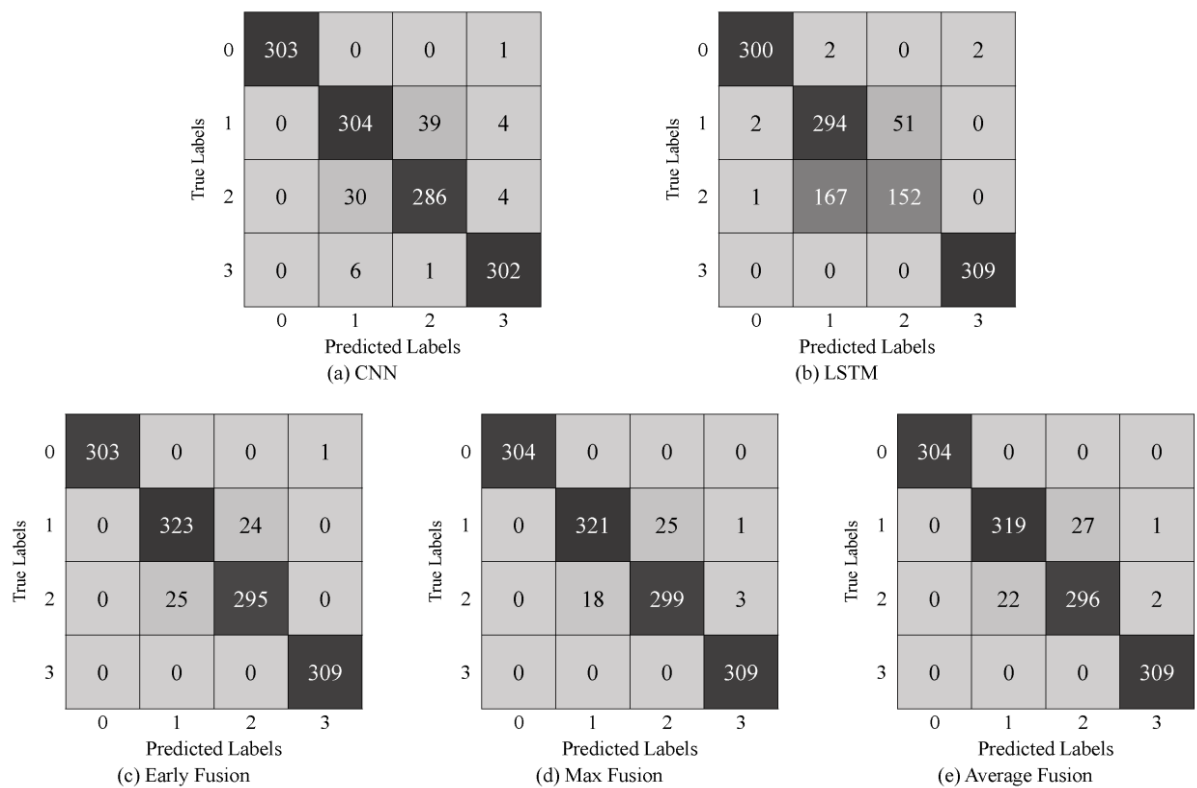

*Figure 9 Confusion matrices (a) CNN, (b) LSTM, (c) Early fusion, (d) Max Fusion, and (e) Average Fusion predictions over a test set of 1280 samples.*

Table 3. Quantitative comparison of the individual models with fused models.

| | Training Accuracy | Testing Accuracy | Class | Precision | Recall | F1 |
|---|---|---|---|---|---|---|
| LSTM Model only | 85% | 82% | No Gas | 0.99 | 0.99 | 0.99 |
| | | | Perfume | 0.63 | 0.85 | 0.73 |
| | | | Smoke | 0.75 | 0.47 | 0.58 |
| | | | Mixture | 0.99 | 1.00 | 1.00 |
| CNN Model only | 95% | 93% | No Gas | 1.00 | 1.00 | 1.00 |
| | | | Perfume | 0.89 | 0.88 | 0.89 |
| | | | Smoke | 0.88 | 0.89 | 0.89 |
| | | | Mixture | 0.97 | 0.98 | 0.97 |
| Early Fusion Model | 97% | 96% | No Gas | 0.1.00 | 1.00 | 1.00 |
| | | | Perfume | 0.93 | 0.93 | 0.93 |
| | | | Smoke | 0.92 | 0.92 | 0.92 |
| | | | Mixture | 1.0 | 1.0 | 1.0 |

The fusion model is trained for 300 epochs, and accuracy and loss curves are analyzed and provides better performance than individual models.

It is evident from the confusion matrices that the false positives and false negatives obtained from the fusion models are considerably lower than the individual models. Hence it can be concluded that the fusion models are outperforming the individual models. Also, the higher testing accuracy of the fusion models demonstrates that the resultant fusion system is more robust and reliable than individual models and performs the task of gas identification and classification with superiority. Analytically, false positives and false negatives appear due to various aspects of the model and data. Primary reason could be the mixing of gases to an extent which makes it difficult for the model to clearly classify. The majority of the false predictions are arriving because of moderate probability of predicting a class. A model can be trained rigorously using the more and varied data samples to solve the false prediction due to boundary line probabilities.

## 5. Conclusions

In this work, a multimodal AI-based fusion framework for reliable identification and detection of gases is developed. We considered four classes (2 individual gases, alcohol vapor obtained from perfume and smoke from incense sticks, 1 as mixture of these gases and 1 no gas) for data collection using sensors, namely, thermal camera for capturing the thermal signature of the gases and array of gas sensors (7 numbers) for detection of specific gases. The data collected is unique and has 5200 samples of both Thermal Images and Gas Sensor Sequence of vector size ($1 \times 7$) Sensors. Both these modalities were fused using Early and Late Fusion Techniques. In summary, the contribution of this work is in bringing in innovative engineering tools for solving a real world problem by developing a more reliable gas detection method involving two modalities and fusing them. The multimodal model outperforms the individual models by supporting or opposing the individual modalities. In case if one modality fails, the other modality can work alone until repair takes place. This is essential in high-risk applications such as leak detection in chemical plants, identification of explosives, etc. The proposed architecture is based on the deep learning frameworks and hence require large number of data samples for appropriate training of the network. The complex data samples involving different combinations of multiple gases in data sample will lead to the robust training of the network. Also, to have efficient and effective operation, a dedicated hardware processing module is essential. The future course of action will focus on the collection of datasets comprising of multiple gases and their combinations in different environmental conditions.

**Author Contributions**

Conceptualization, P.N. and R.W.; methodology, P.N. and R.W.; software, S.M. and P.C.; validation, P.N., R.W., S.M., and P.C.; data curation, S.M. and P.C.; writing—original draft preparation, P.N., R.W., S.M., and P.C.; writing—review and editing, P.N., R.W., K.K., and G.G.; supervision, R.W. and K.K.; funding acquisition, P.N. and R.W. All authors have read and agreed to the published version of the manuscript.

**Funding**

This research was funded by Symbiosis International (Deemed University) under the grant Minor Research Project Grant 2017-18 void letter number SIU/SCRI/MRPAPPROVAL//2018/1769.

# References

1. Trivedi, P.; Purohit, D.; Soju, A.; Tiwari, R.R. Major industrial disasters in India An ofcial newsletter of ENVIS-NIOH, Oct-Dec 2014; Volume 9, No 4. Available online: **http://niohenvis.nic.in/newsletters/vol9_no4_Indian%20Industrial%20Disasters.pdf** (accessed on 9 January 2021).
2. Zhou, Y.; Zhao, X.; Zhao, J.; Chen, D. Research on fire and explosion accidents of oil depots. In Proceedings of the 3rd International Conference on Applied Engineering, Wuhan, China, 22–25 April 2016; Volume 51, pp. 163–168. [**Google Scholar**]
3. Mudur, G.S. Lakhs of early deaths tied to home emissions. *Telegraph India Online* **2015**. Published online on 17 September 2015. Available online: **https://www.telegraphindia.com/india/lakhs-of-early-deaths-tied-to-home-emissions/cid/1513045** (accessed on 15 July 2020).

4. MDC Systems Inc. *Detection Methods. Online Resource*. Available online: **https://mdcsystemsinc.com/detection-methods/** (accessed on 17 July 2020).

5. Fox, A.; Kozar, M.; Steinberg, P. CARBOHYDRATES | Gas Chromatography and Gas Chromatography–Mass Spectrometry. In *Encyclopedia of Separation Science*; Wilson, I.D., Ed.; Academic Press: Cambridge, MA, USA, 2000; pp. 2211–2223. [**Google Scholar**] [**CrossRef**]

6. Wang, T.; Wang, X.; Hong, M. Gas leak location detection based on data fusion with time difference of arrival and energy decay using an ultrasonic sensor array. *Sensors* **2018**, *18*, 2985. [**Google Scholar**] [**CrossRef**] [**PubMed**]

7. Stauffer, E.; Dolan, J.; Newman, R. Chapter 8-Gas Chromatography and Gas Chromatography—Mass Spectrometry. In *Fire Debris Analysis*; Academic Press: Cambridge, MA, USA, 2008; pp. 235–293. [**Google Scholar**] [**CrossRef**]

8. Yin, X.; Zhang, L.; Tian, F.; Zhang, D. Temperature modulated gas sensing E-nose system for low-cost and fast detection. *IEEE Sens. J.* **2015**, *16*, 464–474. [**Google Scholar**] [**CrossRef**]

9. Brahim-Belhouari, S.; Bermak, A.; Shi, M.; Chan, P.C. Fast and robust gas identification system using an integrated gas sensor technology and Gaussian mixture models. *IEEE Sens. J.* **2005**, *5*, 1433–1444. [**Google Scholar**] [**CrossRef**]

10. Majeed, A. Improving time complexity and accuracy of the machine learning algorithms through selection of highly weighted top k features from complex datasets. *Ann. Data Sci.* **2019**, *6*, 599–621. [**Google Scholar**] [**CrossRef**]

11. Khalaf, W.M.H. Electronic Nose System for Safety Monitoring at Refineries. *J. Eng. Sustain. Dev.* **2012**, *16*, 220–228. [**Google Scholar**]

12. Peng, P.; Zhao, X.; Pan, X.; Ye, W. Gas classification using deep convolutional neural networks. *Sensors* **2018**, *18*, 157. [**Google Scholar**] [**CrossRef**]

13. Bilgera, C.; Yamamoto, A.; Sawano, M.; Matsukura, H.; Ishida, H. Application of convolutional long short-term memory neural networks to signals collected from a sensor network for autonomous gas source localization in outdoor environments. *Sensors* **2018**, *18*, 4484. [**Google Scholar**] [**CrossRef**]

14. Pan, X.; Zhang, H.; Ye, W.; Bermak, A.; Zhao, X. A fast and robust gas recognition algorithm based on hybrid convolutional and recurrent neural network. *IEEE Access* **2019**, *7*, 100954–100963. [**Google Scholar**] [**CrossRef**]

15. Liu, Q.; Hu, X.; Ye, M.; Cheng, X.; Li, F. Gas recognition under sensor drift by using deep learning. *Int. J. Intell. Syst.* **2015**, *30*, 907–922. [Google Scholar] [CrossRef]
16. Hamilton, S.; Charalambous, B. *Leak Detection: Technology and Implementation*; IWA Publishing: London, UK, 2020. [Google Scholar]
17. Avila, L.F. Leak Detection with Thermal Imaging. U.S. Patent 6,866,089, 15 March 2005. [Google Scholar]
18. Marathe, S. Leveraging Drone Based Imaging Technology for Pipeline and RoU Monitoring Survey. In *SPE Symposium: Asia Pacific Health, Safety, Security, Environment and Social Responsibility*; Society of Petroleum Engineers: Kuala Lumpur, Malaysia, 2019. [Google Scholar]
19. Jadin, M.S.; Ghazali, K.H. Gas leakage detection using thermal imaging technique. In Proceedings of the 2014 UKSim-AMSS 16th International Conference on Computer Modelling and Simulation, Cambridge, UK, 26–28 March 2014; pp. 302–306. [Google Scholar]
20. Elmenreich, W. A review on system architectures for sensor fusion applications. In *IFIP International Workshop on Software Technolgies for Embedded and Ubiquitous Systems*; Springer: Berlin/Heidelberg, Germany, 2007; pp. 547–559. [Google Scholar]
21. Kalman, R.E. A new approach to linear filtering and prediction problems. *J. Basic Eng. Mar.* **1960**, *82*, 35–45. [Google Scholar] [CrossRef]
22. Luo, Y.; Ye, W.; Zhao, X.; Pan, X.; Cao, Y. Classification of data from electronic nose using gradient tree boosting algorithm. *Sensors* **2017**, *17*, 2376. [Google Scholar] [CrossRef] [PubMed]
23. Simonyan, K.; Zisserman, A. Very deep convolutional networks for large-scale image recognition. *arXiv* **2014**, arXiv:1409.1556. [Google Scholar]
24. Khalaf, W.; Pace, C.; Gaudioso, M. Gas detection via machine learning. *Int. J. Comput. Electr. Autom. Control Inf. Eng.* **2008**, *2*, 61–65. [Google Scholar]
25. Krizhevsky, A.; Sutskever, I.; Hinton, G.E. Imagenet classification with deep convolutional neural networks. *Commun. ACM* **2017**, *60*, 84–90. [Google Scholar] [CrossRef]
26. He, K.; Zhang, X.; Ren, S.; Sun, J. Deep residual learning for image recognition. In Proceedings of the IEEE Conference on Computer Vision and Pattern Recognition, Las Vegas, NV, USA, 27–30 June 2016; pp. 770–778. [Google Scholar]
27. Liu, K.; Li, Y.; Xu, N.; Natarajan, P. Learn to combine modalities in multimodal deep learning. *arXiv* **2018**, arXiv:1805.11730. [Google Scholar]

28. Dong, Y.; Gao, S.; Tao, K.; Liu, J.; Wang, H. Performance evaluation of early and late fusion methods for generic semantics indexing. *Pattern Anal. Appl.* **2014**, *17*, 37–50. [Google Scholar] [CrossRef]
29. Lahat, D.; Adali, T.; Jutten, C. Multimodal data fusion: An overview of methods, challenges, and prospects. *Proc. IEEE* **2015**, *103*, 1449–1477. [Google Scholar] [CrossRef]
30. O'Shea, K.; Nash, R. An introduction to convolutional neural networks. *arXiv* **2015**, arXiv:1511.08458. [Google Scholar]
31. Rumelhart, D.E.; Hinton, G.E.; Williams, R.J. Learning representations by back-propagating errors. *Nature* **1986**, *323*, 533–536. [Google Scholar] [CrossRef]
32. Hochreiter, S.; Schmidhuber, J. Long short-term memory. *Neural Comput.* **1997**, *9*, 1735–1780. [Google Scholar] [CrossRef]
33. Bao, W.; Yue, J.; Rao, Y. A deep learning framework for financial time series using stacked autoencoders and long-short term memory. *PLoS ONE* **2017**, *12*, e0180944. [Google Scholar] [CrossRef] [PubMed]
34. Han, L.; Yu, C.; Xiao, K.; Zhao, X. A new method of mixed gas identification based on a convolutional neural network for time series classification. *Sensors* **2019**, *19*, 1960. [Google Scholar] [CrossRef] [PubMed]
35. Pashami, S.; Lilienthal, A.J.; Trincavelli, M. Detecting changes of a distant gas source with an array of MOX gas sensors. *Sensors* **2012**, *12*, 16404–16419. [Google Scholar] [CrossRef] [PubMed]
36. Awang, Z. Gas sensors: A review. *Sens. Transducers* **2014**, *168*, 61–75. [Google Scholar]
37. Havens, K.J.; Sharp, E.J. *Thermal Imaging Techniques to Survey and Monitor Animals in the Wild: A Methodology*; Academic Press: Cambridge, MA, USA, 2015. [Google Scholar]